\input harvmac
\input epsf

\newcount\figno
\figno=0
\def\fig#1#2#3{
\par\begingroup\parindent=0pt\leftskip=1cm\rightskip=1cm\parindent=0pt
\baselineskip=12pt
\global\advance\figno by 1
\midinsert
\epsfxsize=#3
\centerline{\epsfbox{#2}}
\vskip 14pt

{\bf Fig. \the\figno:} #1\par
\endinsert\endgroup\par
}
\def\figlabel#1{\xdef#1{\the\figno}}
\def\encadremath#1{\vbox{\hrule\hbox{\vrule\kern8pt\vbox{\kern8pt
\hbox{$\displaystyle #1$}\kern8pt}
\kern8pt\vrule}\hrule}}

\overfullrule=0pt

\noblackbox
\parskip=1.5mm
%\def\semi{;~}

%%%%%%%%%%%%%%%%%%%%%%%%%%%%%%%%%%%%%%%%%%%%%%%%%%%%%%%%%%%%%%%%%%%
%%%  modify title page
%%%%%%%%%%%%%%%%%%%%%%%%%%%%%%%%%%%%%%%%%%%%%%%%%%%%%%%%%%%%%%%%%%%
\def\Title#1#2{\rightline{#1}\ifx\answ\bigans\nopagenumbers\pageno0
%   \vskip0.5in
\else\pageno1\vskip.5in\fi \centerline{\titlefont #2}\vskip .3in}

\font\caps=cmcsc10
%\def\listrefs{\footatend\bigskip\bigskip\immediate\closeout\rfile
%\writestoppt \baselineskip =11pt\centerline{{\secfont References}}
%\bigskip{\frenchspacing\parindent =20pt \escapechar +'
%\input\jobname.refs \vfill\eject}\nonfrenchspacing} 
%%%%%%%%%%%%%%%%%%%%%%%%%%%%%%%%%%%%%%%%%%%%%%%%%%%%%%%%%%%%%%%%%%%%%%%%%%%%

\noblackbox
\parskip=1.5mm
%\def\semi{;~}

%%%%%%%%%%%%%%%%%%%%%%%%%%%%%%%%%%%%%%%%%%%%%%%%%%%%%%%%%%%%%%%%%%%%%
  
\def\npb#1#2#3{{\it Nucl. Phys.} {\bf B#1} (#2) #3 }
\def\plb#1#2#3{{\it Phys. Lett.} {\bf B#1} (#2) #3 }
\def\prd#1#2#3{{\it Phys. Rev. } {\bf D#1} (#2) #3 }
\def\prl#1#2#3{{\it Phys. Rev. Lett.} {\bf #1} (#2) #3 }
\def\mpla#1#2#3{{\it Mod. Phys. Lett.} {\bf A#1} (#2) #3 }

\def\cmp#1#2#3{{\it Commun. Math. Phys.} {\bf #1} (#2) #3 }

\def\bb#1{{\tt hep-th/#1}}

\def\jhep#1#2#3{{\it J. High Energy Phys.} {\bf #1} (#2) #3 }

%%%%%%%%%%%%%%%%%%%%%%%%%%%%%%%%%%%%%%%%%%%%%%%%%%%%%%%%%%%%%%%%%%%%%
%%%%%%%%%%%%%%%%%%%%    some definitions    %%%%%%%%%%%%%%%%%%%%%%%%%
%%%%%%%%%%%%%%%%%%%%%%%%%%%%%%%%%%%%%%%%%%%%%%%%%%%%%%%%%%%%%%%%%%%%%

           \def\CO{{\cal O}} 
   
\def\CL{{\cal L}}   
 \def\CR{{\cal R}}  
 
\def\CN{{\cal N}}

%%%%%%%%%%%%%%%%%%%%%%%%%%%%%%%%%%%%%%%%%%%%%%%%%%%%%%%%%%%%%%%%%%%%%

\def\dj{\hbox{d\kern-0.347em \vrule width 0.3em height 1.252ex depth
-1.21ex \kern 0.051em}}

\def\pt{\partial}

\def\Dirac{\,\raise.15ex\hbox{/}\mkern-13.5mu D}
\def\dirac{\,\raise.15ex\hbox{/}\kern-.57em \partial}
\def\shalf{{\ifinner {\textstyle {1 \over 2}}\else {1 \over 2} \fi}} 
\def\sshalf{{\ifinner {\scriptstyle {1 \over 2}}\else {1 \over 2} \fi}} 
\def\sfourth{{\ifinner {\textstyle {1 \over 4}}\else {1 \over 4} \fi}}
%%%%%%%%%%%%%%%%%%%%%%%%%%%%%%%%%%%%%%%%%%%%%%%%%%%%%%%%%%%%%%%%%%%%%%
%%%%%%%%%%%%%%%%%%%%%%%        references         %%%%%%%%%%%%%%%%%%%%%%
%%%%%%%%%%%%%%%%%%%%%%%%%%%%%%%%%%%%%%%%%%%%%%%%%%%%%%%%%%%%%%%%%%%%%%%
%%%%%%%%%%%%%%%%%%%
%%%

\lref\rbek{J.D. Bekenstein, \prd{7}{1973}{2333.} \prd{9}{1974}{3292.}
\prd{12}{1975}{3077.}}

\lref\rhaw{S.W. Hawking, {\it Commun. Math. Phys.} {\bf 43} (1975) 199.
\prd{13}{1976}{191.}}

\lref\rgibbhaw{G.W. Gibbons and S.W. Hawking, \prd{15}{1977}{2752.}}

\lref\raps{A. Adams, J. Polchinski and E. Silverstein, \bb{0108075.}}

\lref\rHP{G.T. Horowitz and J. Polchinski, \prd{55}{1997}{6189,}
\bb{9612146.}}

\lref\rholo{G. 't Hooft, {\tt gr-qc/9310026.} J.D Bekenstein, \prd{49}{1994}{1912.}
L. Susskind, {\it J. Math. Phys.} {\bf 36} (1995) 6377, \bb{9409089.}}

\lref\rkog{I.I. Kogan, {\it Sov. Phys. JETP Lett.} {\bf 45} (1987) 709. 
B. Sathiapalan, \prd{35}{1987}{3277.}}

\lref\raw{J.J. Atick and E. Witten, \npb{310}{1988}{291.}}

\lref\rads{J. Maldacena, {\it Adv. Theor. Math. Phys.} {\bf 2} (1998)
231 \bb{9711200.}
 S.S. Gubser, I.R. Klebanov and A.M. Polyakov,
\plb{428}{1998}{105} \bb{9802109.} E. Witten,
{\it Adv. Theor. Math. Phys.} {\bf 2} (1998)
253 \bb{9802150.} }

\lref\rbunch{M.S. Costa and M. Gutperle, \jhep{03}{2001}{027} \bb{0012072.}
 M. Gutperle
and A. Strominger, \jhep{06}{2001}{035} \bb{0104136.} }

\lref\rstrvaf{A. Strominger and C. Vafa, \plb{379}{1996}{99} \bb{9601029.}} 

\lref\rwitbub{E. Witten, \npb{195}{1982}{481.} } 

\lref\rsen{A. Sen, \bb{9904207.}}

\lref\rmaldacovi{N. Itzhaki, J.M. Maldacena, J. Sonnenschein and S. 
Yankielowicz, \prd{58}{1998}{046004} \bb{9802042.}}

\lref\rbdm{T. Banks, M.R. Douglas, G.T. Horowitz and E. Martinec,
\bb{9808016.}}

\lref\rdabho{A. Dabholkar, \bb{0111004.}}

\lref\rchicago{J.A. Harvey, D. Kutasov, E.J. Martinec and G. Moore,
\bb{0111154.}}

\lref\rgutperlem{J.R. David, M. Gutperle, M. Headrick and S. Minwalla,
\bb{0111212.}}

\lref\rgutstr{M. Gutperle and A. Strominger, \jhep{0106}{2001}{035} 
\bb{0104136.}} 

\lref\rhag{R. Hagedorn, {\it Supp. Nuovo Cim.} {\bf 3} (1965) 147.} 

\lref\rhist{S. Fubini and G. Veneziano, {\it Nuovo Cim.} {\bf 64A} (1969)
1640. K. Huang and S. Weinberg, \prd{25}{1970}{895.}
 S. Frautschi, \prd{3}{1971}{2821.} R.D. Carlitz,
\prd{5}{1972}{3231.}
E. Alvarez, \prd{31}{1985}{418;} \npb{269}{1986}{596.}
M. Bowick and L.C.R. Wijewardhana, \prl{54}{1985}{2485.} B. Sundborg,
\npb{254}{1985}{883.} S. N. Tye, \plb{158}{1985}{388.} E. Alvarez and
M.A.R. Osorio, \prd{36}{1987}{1175.} P. Salomonson and B. Skagerstam,
\npb{268}{1986}{349.} D. Mitchell and N. Turok, \prl{58}{1987}{1577.}
}

\lref\rkogan{I.I. Kogan, {\it JETP. Lett.} {\bf 45} (1987) 709.
 B. Sathiapalan, \prd{35}{1987}
{3277.}}

\lref\rhh{S.W. Hawking and G. Horowitz, {\it Class. Quant. Grav.} {\bf 13} 
(1996) 1487, {\tt gr-qc/9501014.}}

\lref\rstretch{L. Susskind, \bb{9309145.} L. Susskind and J. Uglum,
\prd{50}{1994} \bb{9401070.}
A. Sen,  {\it Mod. Phys. Lett. }{\bf A10} (1995) 2081 
\bb{9504147.}}

\lref\rbmth{G. Horowitz and L. Susskind, {\it J. Math. Phys.} {\bf 42} (2001)
3158 \bb{0012037.} S.P. de Alwis and A.T. Flournoy, \bb{0201185.}
R. Emparan and M. Guperle, \jhep{0112}{2001}{023}  
\bb{0111177.} M. Gutperle, \bb{0207131.}}

\lref\rforge{S. Elitzur, A. Forge and E. Rabinovici, \npb{359}{1991}{581.}}

\lref\rgkleb{D.J. Gross and I.R. Klebanov, \npb{344}{1990}{475.}}

\lref\rushag{J.L.F. Barb\'on and E. Rabinovici, \jhep{03}{2002}{057} \bb{0112173.}} 

\lref\rcmalda{C.G. Callan and J.M. Maldacena, \bb{9708147.}}

\lref\rcorrh{M. Bowick, L. Smolin and L.C.R. Wijewardhana, {\it Gen. Rel.
Grav.} {\bf 19} (1987) 113. G. Veneziano, {\it Europhys. Lett.}
 {\bf 2} (1986) 199.
L. Susskind, \bb{9309145.} G. Veneziano, in {\it Hot Hadronic Matter:
Theory and Experiments}, Divonne, June 1994, eds J. Letessier, H. Gutbrod
and J. Rafelsky, NATO-ASI Series B: Physics, {\bf 346} (1995), p. 63.
A. Sen, \mpla{10}{1995}{2081.} E. Halyo, A. Rajaraman and L. Susskind,
\plb{382}{1997}{319,}
\bb{9605112.} E. Halyo, A. Rajaraman, B. Kol and L. Susskind,
\plb{401}{1997}{15,}
\bb{9609075.}}

\lref\rpolyakov {A.M. Polyakov, {\it 
``Gauge Fields and Strings"}, Harwood Academic
Publishers (1987). Page 265.} 

\lref\rpolbook {J. Polchinski, {\it ``String Theory"},
 Volume II, Cambridge University
Press (1998).}  

\lref\rusabel {S.A. Abel, J.L.F. Barb\'on, I.I. Kogan and
 E. Rabinovici,
\jhep{9904}{1999}{015,}
\bb{9902058.} See also the contribution to {\it ``Many Faces of the Superworld"}:
 Yuri Golfand Memorial Volume, World Scientific, Singapore (1999) \bb{9911004.}}

\lref\rusex{J.L.F. Barb\'on and E. Rabinovici, \npb{545}{1999}{371} \bb{9805143.} 
J.L.F. Barb\'on, I.I. Kogan and E. Rabinovici,
\npb{544}{1999}{104} \bb{9809033.}}

\lref\rbooks{M.B. Green, J.H. Schwarz and E. Witten, {\it ``Superstring
Theory"}, Vols I, II. Cambridge Monographs on Mathematical Physics
(1987). J. Polchinski, {\it ``String Theory"}, Vols. I, II. Cambridge
University Press (1998).} 

\lref\rers{S. Elitzur, E. Rabinovici and G. Sarkisian, \npb{541}{1999}{246}
\bb{9807161.}} 

\lref\rmarti{E.J. Martinec, \bb{0210231.}} 

\lref\rzamo{A.A. Zamolodchikov, {\it JETP Lett.} {\bf 43} (1986) 730.}

\lref\rbath{O. Aharony, M. Fabinger,  G.T. Horowitz and E. Silverstein,  
\jhep{0207}{2002}{007}  
\bb{0204158.}}

\lref\rHPage{S.W. Hawking and D. Page, \cmp{87}{1983}{577.}}

\lref\rgpy{D.J. Gross, M.J. Perry and L.G. Yaffe, \prd{25}{1982}{330.}}

\lref\rwithp{E. Witten, {\it Adv. Theor. Math. Phys.} {\bf 2} (1998)
505 \bb{9803131.}}

\lref\ralud{For boundary flows, see I. Affleck and A.W. Ludwig, \prl{67}{1991}{161.}}

%%%%%%%%TEXT%%%%%%%%%%%%%%%%%%%%%%%%%%%%%%%%%%%%%%%%%%%%%%%%%%%%%%%%%
%%%%%%%%%%%%%%%%%%%%%%%%%%%%%%%%%%%%%%%%%%%%%%%%%%%%%%%%%%%%%%%%%%%%%
%%%%%%%%%%%%%%%%%%          title page       %%%%%%%%%%%%%%%%%%%%%%%%%
%%%%%%%%%%%%%%%%%%%%%%%%%%%%%%%%%%%%%%%%%%%%%%%%%%%%%%%%%%%%%%%%%%%%%%

\baselineskip=15pt

\line{\hfill CERN-TH/2002-313}
%\line{\hfill DFTT 15/2000}
%\line{\hfill SPIN-2000/13}
%\line{\hfill UB-ECM-PF-00/06}
%\line{\hfill ITFA-2000-08}
\line{\hfill {\tt hep-th/0211212}}

\vskip 0.5cm

\Title{\vbox{\baselineskip 12pt\hbox{}
 }}
{\vbox {\centerline{Remarks on Black Hole Instabilities  }
\vskip10pt
\centerline{and Closed String Tachyons
}
}}

\vskip1.2cm

\centerline{$\quad$ {\caps 
J.L.F. Barb\'on~$^{a,}$\foot{ On leave
from Departamento de F\'{\i}sica de Part\'{\i}culas da 
Universidade de Santiago de Compostela, Spain.} and 
E. Rabinovici~$^{b}$ 
}}
\vskip0.5cm

\centerline{{\sl $^a$ Theory Division, CERN, 
 CH-1211 Geneva 23, Switzerland}}
\centerline{{\tt 
barbon@cern.ch}} 

\vskip0.2cm

\centerline{{\sl $^b$ Racah Institute of Physics, The Hebrew University 
 Jerusalem 91904, Israel}}
\centerline{{\tt eliezer@vms.huji.ac.il}}

\vskip1.5cm

\centerline{\bf ABSTRACT}

 \vskip 0.1cm

 \noindent 

 Physical arguments stemming from the theory of black-hole thermodynamics
are used to put constraints on the dynamics of closed-string tachyon  
condensation in Scherk--Schwarz compactifications. 
 A  geometrical interpretation of the tachyon condensation involves 
 an effective capping of a noncontractible cycle, thus removing the
very topology that supports the tachyons. A semiclassical regime
is identified in which the matching between the tachyon condensation and
the black-hole instability flow is possible. We formulate a generalized
correspondence principle and illustrate it in several different circumstances: 
an Euclidean interpretation of the transition from strings to black holes
across the Hagedorn temperature and instabilities in the brane-antibrane
system. 
\foot{Contribution to Jacob Bekenstein's Festschrift. }

\vskip 0.1cm

\Date{October  2002}
               
%\noindent CERN-TH/2001-*** 

\vfill

%\vskip 0.1cm 

%%%%%%%%%%%%%%%%%%%%%%%%%%%%%%%%%%%%%%%%%%%%%%%%%%%%%%%%%%%%%%%%%%%%%%

%\draft

%%%%%%%%%%%%%%%%%%%%%%%%%%%%%%%%%%%%%%%%%%%%%%%%%%%%%%%%%%%%%%%%%%%%%%%%%%
%%%%%%%%%%%%                text begins                        %%%%%%%%%%%
%%%%%%%%%%%%%%%%%%%%%%%%%%%%%%%%%%%%%%%%%%%%%%%%%%%%%%%%%%%%%%%%%%%%%%%%%%

\baselineskip=14pt

\newsec{Introduction}

\noindent

The relation between string theory and gravity was not one of love at
first sight.  String theory was born in
an effort to explain the strong interactions and still treats them
as an old flame.
In the original open string formulation the effective low
energy theory was that of a gauge theory. 
Unitarity however, required the
existence of a closed string sector as well.
The low energy theory of closed strings included gravity \refs\rbooks.
Once it was realized that gravity is involved string theory attempted to
generalize and replace general relativity,  this did not serve to endear
string theory to some practioners of general relativity.

In some sense, a full circle was closed with the nonperturbative formulation
of some closed string theories
 in terms of large-$N$ gauge theories living on the boundary
of spacetime \refs\rads.  In these developments, one often finds  string
theory looking  
 for  guidance into, general relativity, the very framework it would
wish to generalize.

In this contribution we will describe several instances of such ``retarded impact"
of general relativity in string theory. 
 One such case involves attempts to construct phase diagrams of gravity.
String theory is a theory in which, at least it seems, the basic
constituents are extended objects. This has many ramifications on the
symmetries and the dynamics of the theory.
One of the very striking new features of strings is their very large
entropy; for free strings it is a  linear function of the energy. For a
regular field theory in $d$ spatial dimensions the entropy only increases as
$E^{d/d+1}$. One consequence of
the large string entropy,  if it persists for all energies and for
finite couplings, is that the system of strings
has a maximal temperature called the Hagedorn temperature. The Hagedorn
temperature is one of the stop signs appearing in string theory, which
may  suggest
for example an upper bound to the temperature of the universe at short
time scales.  In the strong interactions  setting, it was suggested
that this temperature, rather than being limiting, was
signaling the limits of the
effective description of the theory at that temperature. The constituents
of QCD field theory are quarks and gluons, not hadrons.  They manifest themselves
clearly at  high energies and temperatures. 

One can address this question also in closed string theory.
 Can one cross the Hagedorn temperature? Will one uncover more basic
constituents of strings in the  process?
Here general relativity offered some guidance and some resolutions.
The entropy of black holes \refs{\rbek, \rhaw}\ and the property of holography
\refs\rholo, both uncovered
by Bekenstein, play a key role in the attempts to understand the
significance of the Hagedorn temperature.
It turns out that when there is a more basic picture of constituents
of strings such as in the AdS/CFT correspondence case, the Hagedorn
temperature can be surpassed; in other cases the Hagedorn
temperature remains the maximal temperature even after black holes are
formed. In both cases black holes and their entropy dominate  at the high energy
scale.
One may draw phase diagrams of string/gravitational systems by
constructing a cocktail of, among others, massless fields, strings,
branes, black branes and various types of black holes. Such phase diagrams,
constructed along the blueprints of \refs\rHP\ and \refs\rmaldacovi, 
generally are consistent and complete. We present the simplest  
example in Figure 1 (see, for instance \refs{\rbdm, \rusabel}). This is the
phase diagram of type IIB string theory on ${\rm AdS}_5 \times {\bf S}^5$
with $N$ units of RR flux. Alternatively, it is also the phase diagram of
$\CN=4$ Super Yang--Mills theory on ${\bf S}^3$. The different regions
in this picture are labelled by the degrees of freedom that dominate the
density of states.

\fig{\sl The phase diagram of type IIB strings on ${\rm AdS}_5 \times {\bf S}^5$
with $N$ units of RR flux, as a function of the string coupling $g_s$ and
the  energy, in units of the curvature radius $R$. 
 The phase diagram can be
continued past $g_s =1$  using the self-duality of the theory under
$g_s \rightarrow 1/g_s$.    
}{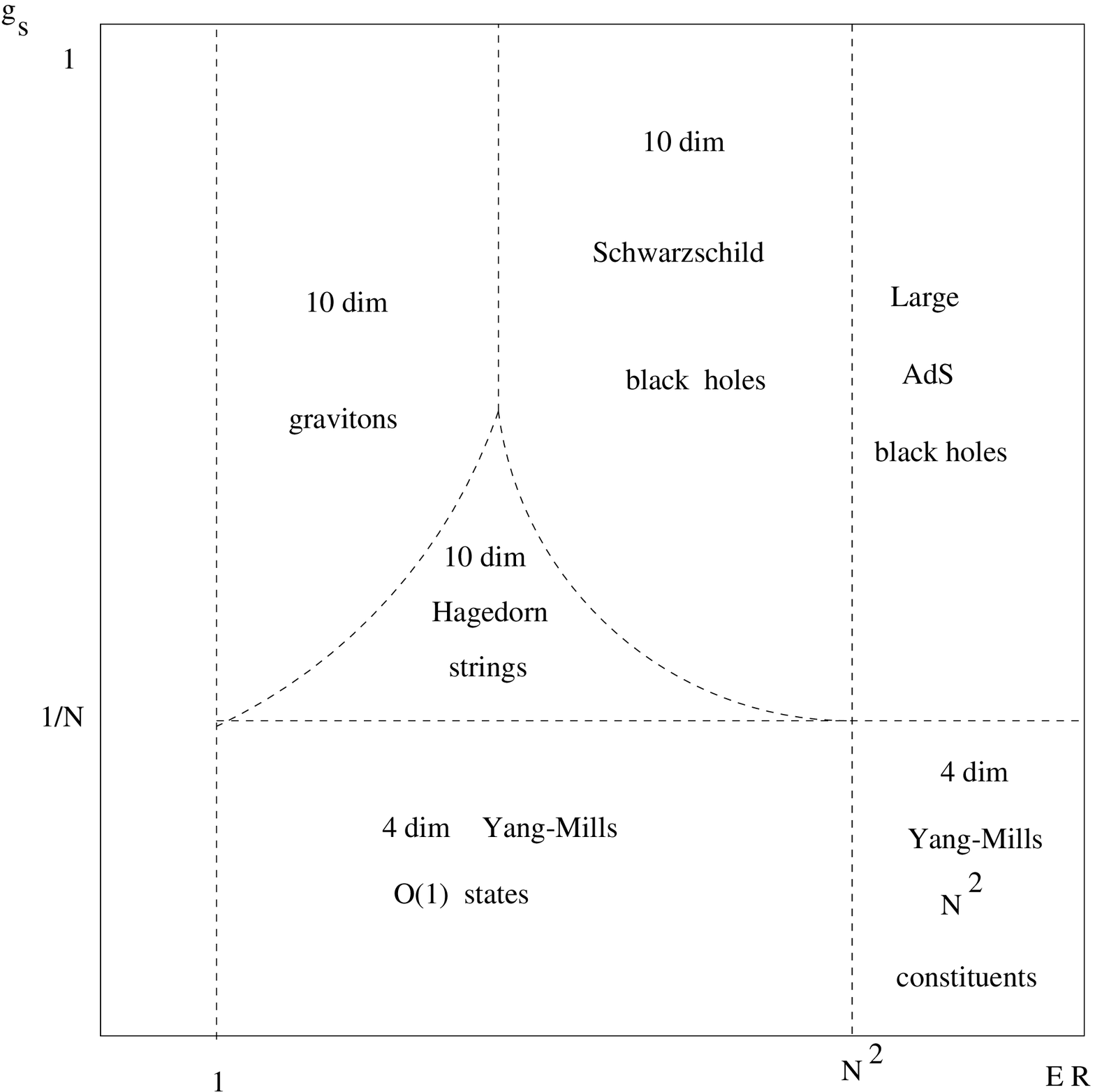}{5truein}

If the phase diagram is plotted as a function of temperature rather than energy
(i.e. going to the canonical rather than the microcanonical ensemble) one finds
that all phases dominated by
 {\it localized} degrees of freedom in ten dimensions
dissapear (gravitons, strings and Schwarzschild black holes). This shows that
these configurations are unstable at fixed temperature. The canonical instability
of Schwarzschild black holes is well known on account of their negative specific
heat. A thermal gas of gravitons  is also
 unstable to gravitational collapse through the Jeans
instability. Both processes end up with a 
large Anti-de Sitter  black hole, which
is the only stable configuration at large temperatures. Since such black holes 
are larger than the curvature radius of AdS, 
  all scales that look approximately
like flat ten-dimensional space disappear behind the horizon.

Thus, it is natural to connect the remaining unstable phase, that of Hagedorn
strings in ten dimensions, to an instability that ends in the large AdS black hole.   
At a technical level, the Hagedorn instability is related to the condensation of
tachyonic winding modes, when studied in the canonical Euclidean formalism
 \refs{\rkog, \raw}. Hence, Euclidean 
black-hole dynamics is bound to teach us some general
 lessons about
tachyon condensation processes in string theory.   This
will be the main theme of  this contribution.

String theory has been known from its inception to generically contain
perturbative 
instabilities, that is tachyons. In special circumstances these tachyons
are absent in the physical sector. Both bulk and boundary tachyons were
studied as relevant operators in the CFT
worldsheet theory. 
 Following their flow  one needs to
understand how to maintain fixed the value of the Virasoro central charge.
For boundary tachyons the value remains unchanged during the flow, whereas for
bulk tachyons it seems that large parts of spacetime disappear during the
flow (for boundary tachyons only small parts may disappear). In most cases
it is very difficult to follow the flow to a new vacuum. 

Euclidean black holes also tend to gobble-up chunks of
spacetime as they evolve. Moreover, their topology can be related to
their unique thermodynamical properties. We  observe 
 that certain instabilities of near-extremal black holes should
be identified with tachyonic instabilities in Scherk--Schwarz 
compactifications.  
 We suggest to use
general
thermodynamical properties of black holes to identify and constrain the
possible flows of stringy tachyon instabilities. This should be done in
the regions of the flow which are reliably treated semiclassically.
The common feature of the cases treated is that the tachyons can
form in
the stringy regime because a nontrivial 
topology supports them. We suggest  that, 
upon condensation, a topology change occurs that stabilizes the
system, in such a way that the tachyons  undo
the very topology that enabled their appearence.
 The overall geometrical picture bears a  resemblance
to recent discussions of tachyon condensation in orbifolds (see
for example \refs\rmarti\ for a recent review). 

The paper is organized as follows. In Section 2 we 
 review some standard results and some less known subtleties
 on the relation between  black-hole entropy
and spacetime topology. 
In Section 3 we consider warped Scherk--Schwarz compactifications
as our model of tachyon instability and we formulate our 
hypothesis  about  the geometrical interpretation of the corresponding
flows.  In Section 4 we consider two examples of the same phenomenon
in which the physics is under better control. This is the example
of the Hagedorn transition in AdS space, and the semiclassical
picture of D-brane/antiD-brane annihilation.   In Section 5 we
abstract from these examples 
 a generalization of the string/black-hole correspondence
principle \refs{\rHP, \rcorrh}\ that applies to off-shell effective actions.

\newsec{Black Hole Entropy and Spacetime Topology}

\noindent

One of the most significant aspects of the Bekenstein--Hawking entropy
formula \refs{\rbek, \rhaw} 
 is its mysterious   relation to spacetime topology \refs\rgibbhaw.
In the semiclassical approximation to the  Euclidean functional integral,
 asymptotic boundary conditions for a thermal ensemble
at temperature $T=1/\beta$ are such that the metric at infinity must
approach ${\bf R}^3 \times {\bf S}^1$ with ${\rm Length}\,({\bf S}^1) = \beta$.
A Schwarzschild
black hole  in (unstable) equilibrium with the thermal ensemble is described by
the classical geometry
\eqn\sbh{
ds^2 = \left(1-{2GM \over r} \right) \,d\tau^2 + \left(1-{2GM \over r} 
\right)^{-1}
 \;dr^2  
+ r^2 \,d\Omega^2
\,,}
with the identification  $\tau \equiv \tau + \beta$. The Hawking temperature
$1/\beta$ is related  to the ADM mass by 
 $\beta= 8\pi  GM$, in order to avoid the conical singularity at
$r=2GM$.

The Euclidean gravitational  action
of a manifold $X$ is given by
\eqn\ega{
I(X)= -{1\over 16\pi G } \int_X \;(\CR - 2\Lambda) - {1\over 8\pi G}
 \oint_{\pt X} \;{\cal K} +
C_{\infty}
\,,} 
with $C_\infty$ an  appropriate set of   counterterms at non-compact
boundaries. When this action is
evaluated on the Euclidean Schwarzschild solution, $I(X)$ 
 is interpreted as the canonical free
energy. Now, we would expect any internal degrees of freedom of the
black hole to show up at the one-loop level. At the classical level
we may imagine that the result should be that of a heavy static particle
of mass $M$, that is $I = \beta\,M$. However, explicit
calculation yields half of this result, i.e. one finds $I = \beta M/2$.
This we can rewrite as
$
I= \beta\,F= \beta\,M -S
$, 
with $S$ representing an effective ``classical" entropy. The result is
\eqn\bhen{
S= {A_H \over 4G} = 4\pi\,G\, M^2
\;,} 
the Bekenstein--Hawking entropy.
Therefore, the formalism ``mimics" the fact that black holes have the microstate
degeneracy envisaged by Bekenstein, even if the entropy appears as a purely
classical feature in this geometry. 

In fact, this classical contribution to the entropy has a topological  
interpretation. Notice that the Schwarzschild manifold \sbh\   
is simply connected, with  topology ${\bf R}^2 \times {\bf S}^2$. This is unlike
the  standard topology associated to thermal boundary conditions, ${\bf S}^1 \times
{\bf R}^3$, which has  a  non-contractible circle.
  We can decompose the Euclidean Schwarzschild section
as $X_\epsilon  \,\cup \, (X-X_\epsilon)$, where
$X_\epsilon = D_\epsilon \times {\bf S}^2$  and $D_\epsilon$ is
a microscopic disc of radius $\epsilon$ cut out around $r=2GM$.
  The piece $X-X_\epsilon$ has cylindrical  topology and
gives the standard Hamiltonian result:
$$
\lim_{\epsilon \to 0} \;I(X-X_\epsilon) = \beta\,M
\;,
$$
whereas the  non-Hamiltonian part, $X_\epsilon$, 
 gives the entropy:
$$
\lim_{\epsilon \to 0} \;I(X_\epsilon) = -S
\;.$$

The {\it ab initio} statistical interpretation of
this ``classical" entropy has been a long-standing problem that was
finally solved for models with holographic dual interpretation \refs\rads. In
these cases the classical gravity approximation emerges as the
large-$N$ master field of some dual gauge theory living on the
boundary of spacetime. In this sense, the classical gravitational
entropy computes
the planar approximation to the {\it quantum} entropy of the dual theory.

This topological avatar  of the black-hole entropy leads to an interesting
puzzle in the case of extremal black holes. 
Consider a Reissner--Nordstrom (RN) black hole with charge $Q<M$ and
Euclidean metric  
\eqn\rnbh{
ds^2 =\left(1-{2M \over r} + { Q^2 \over r^2}\right) \,d\tau^2 +
\left(1-{2M \over r} + {Q^2 \over r^2}\right)^{-1} \;dr^2 + r^2 \;d\Omega^2
\;,} 
where we have switched to $G=1$ units.
  Now the Hawking inverse temperature is  related to $M$ and $Q$ by
\eqn\hh{
\beta= {4\pi\,r_0^3 \over r_0^2 -  Q^2}
\;,}
where the horizon radius is 
\eqn\hrad{
 r_0 (M) = M + \sqrt{M^2 - Q^2}
\;.}
Notice that the function $\beta(M,Q)$ diverges as $(M-Q)^{-1/2}$ in the 
 extremal limit $M\rightarrow Q$, i.e. the extremal RN black hole has
zero temperature. On the other hand, we recover the Schwarzschild form for
$M\gg Q$. For a given fixed charge $Q$, there is a minimum 
$
\beta_{\rm min} = 6\pi \sqrt{3}\,Q  
$
 occurring at $M_c = Q/\sqrt{3}$. This means that, for a fixed $Q$,
RN black holes have a maximal temperature $T_{\rm max} = 1/\beta_{\rm min}$.
For temperatures below the maximal,
 there are two
RN black holes with the same temperature, one with $M=M_+ (\beta) >M_c$
 that has negative
specific
 heat, and another one  
with $M=M_- (\beta) <M_c$ that is thermodynamically stable (it 
  has positive specific heat).

For  extremal black holes with  $Q=M$ the horizon at $r=Q$ degenerates and
sits at the end of a ``throat" with infinite proper length.
The resulting Euclidean  geometry
 has cylindrical topology ${\bf S}^1 \times {\bf R}^3$, i.e. it
is a ``Hamiltonian" thermal manifold.  
 Therefore,  despite
the fact that the extremal black hole has non-zero horizon area,
one could imagine assigning 
 to it a zero entropy since $I=\beta\,M$ in this case \refs\rhh. This is at
odds with modern microscopic determinations of the entropy for extremal
supersymmetric black holes
that can be embedded in string theory \refs\rstrvaf.
 In these cases, a non-zero horizon
area is always associated to a true degeneracy of supersymmetric ground states
of the system. Hence the puzzle. 

In fact, the definition that assigns non-zero entropy to the extremal black hole
turns out to be more physical.   
  The asymptotic boundary conditions
of the $M=Q$ Euclidean metric imply a non-vanishing temperature, whereas
the intrinsic temperature of the extremal black hole is zero. Physically, this
situation corresponds to introducing a zero-temperature extremal black hole
into a thermal bath. The black hole will acrete energy from the bath, thereby
increasing its mass, until it matches that of the stable black hole
with temperature $1/\beta$ and mass $M_- (\beta)$.  Therefore, stable configurations
have a non-vanishing limiting entropy as $M\rightarrow Q$.   

In the Euclidean formalism, we can approximate the black-hole
 growth process by a set of interpolating metrics, 
$X_M$, with effective mass in the  range $Q< M < M_- (\beta)$, and
with fixed thermal boundary conditions (temperature) at infinity.
 Such metrics have
the form \rnbh\ with a conical singularity at $r=r_0$, which is still given by
\hrad. However, the asymptotic temperature does 
not satisfy \hh\  and the interpolating
metrics are not
 solutions of Einstein's equations (due to the conical singularity).
 Along the interpolating
family, the Euclidean gravitational 
action 
$$
I(X_M) = \beta\,M - S=  \beta\,M - \pi\,r_0^2 (M) 
$$ 
 decreases monotonically between the initial
and final true solutions (c.f. Fig 2). This is exactly the expected behaviour
for a thermal effective potential. 

\fig{\sl The Euclidean action of the interpolating black-hole metrics with fixed
$\beta > \beta_{\rm min}$.
 There is a local minimum at the stable 
near-extremal black hole $M=M_- (\beta)$ 
and a local maximum at the unstable RN
 Schwarzschild-like black hole $M_+ (\beta)$.
}{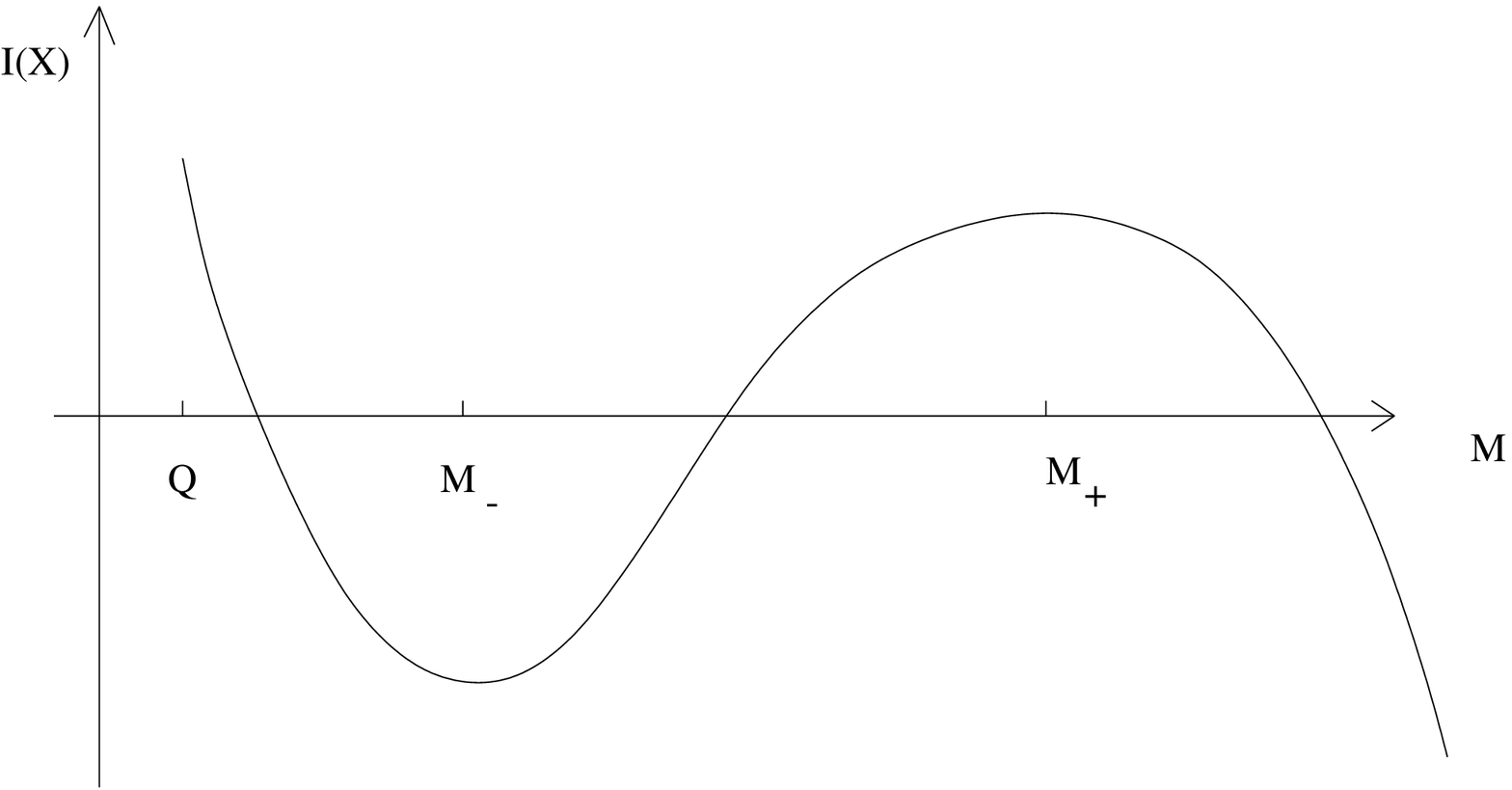}{5truein}

If the RN metric is embedded in string theory, what possible interpretation
could the off-shell  interpolating metrics have? 
 One
context in which off-shell spacetime manifolds have a meaning is 
in terms of world-sheet renormalization-group (RG) flows. 
World-sheet conformal fixed points correspond to classical
solutions to the background equations of motion, while departures from
conformality can be viewed as off-shell deformations. With this analogy
in mind, we will refer
to the set of interpolating  metrics as a ``semiclassical flow".  

\newsec{Closed-String Tachyons and Off-Shell Topology Change}

\noindent

Closed-string tachyon condensation is a difficult dynamical problem, partly
because it involves background dynamics in the infrared, i.e. the spacetime
at infinity is unstable, even at string tree level.
For tachyons that are localized at defects or ``impurities", the   
analysis of the dynamics  is more succesful. 
Such is the flow in open string theories \refs\rers, the decay
of D-branes 
 \refs\rsen, or tachyonic orbifolds 
 \refs{\raps}.  
 This suggests an attempt at ``quasilocalizing" the closed-string tachyons.
One  possibility is to use tachyonic winding modes from a {\it warped}
Scherk--Schwarz compactification. Consider backgrounds
 with topology $X={\bf S}^1
\times Y$ where the proper length of the circle $L({\bf S}^1)$ varies
monotonically  as a
function of some  ``radial" direction of $Y$.    
Let us choose the radial coordinate so that $L({\bf S}^1)_r$ increases
with $r$ and define a ``correspondence radius" $r_s$  by 
\eqn\corad{
L\,({\bf S}^1)_{r=r_s} = \ell_s\,
,}
where $\ell_s \sim \sqrt{\alpha'}$ is the fundamental string length
scale. 
Assuming adiabaticity in $r$, 
light winding modes will be supported  for $r<r_s$, since $L({\bf S}^1 
)<\ell_s$ in that region. Moreover, if ${\bf S}^1$ is endowed with a 
supersymmetry-breaking spin structure, these light modes will be
generically tachyonic, 
 due to the negative contribution of the
world-sheet Casimir energy.

As a result, the region $r<r_s$ acts like a ``box" that localizes the
tachyons, providing an infrared cutoff. This is analogous to the 
twisted tachyons in orbifolds, which are nothing but winding modes
around the tip of the cone. The orbifold example also shows that
tachyons are generic when supersymmetry is broken, even if the
adiabatic approximation is violated.  On the other hand, depending
on the geometry of $Y$, we  
might be able to tune the size of the box and make it macroscopic
in string units. In this way we can study the condensation of
``quasilocalized" tachyons. 

\fig{\sl Adiabatic picture of the tachyonic Scherk--Schwarz compactification.
Tachyonic winding modes are confined to $r<r_s$. The RG flow ends with
the ``capping of the throat" at $r=r_0 (M_-)$
 and the removal of the nontrivial topology that supported the
tachyons.}  
{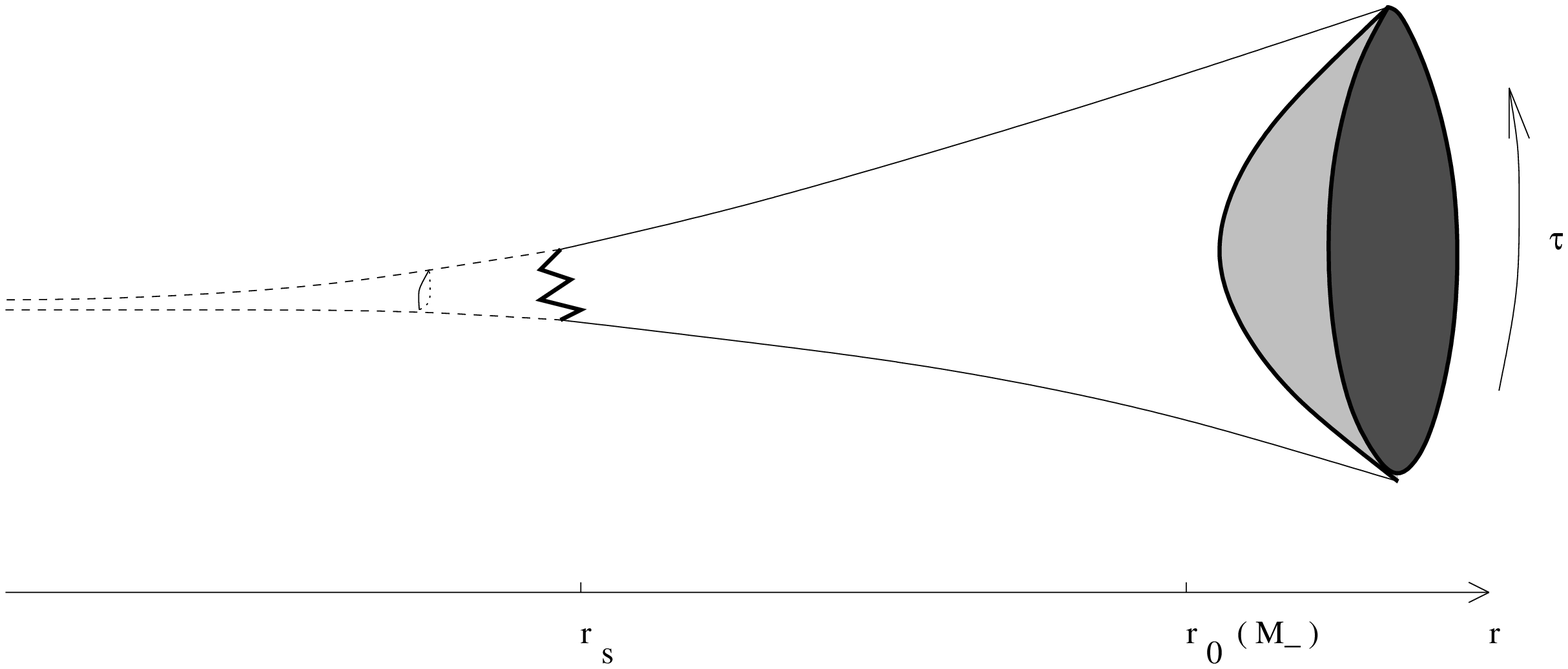}{5truein}

If the dilaton can be kept small throughout the background $X$,
 the process of closed-string tachyon condensation can be viewed
as some RG flow on the world-sheet, or
 as a flow in the space of off-shell string
backgrounds.   
This set up is qualitatively similar to our  discussion of
off-shell  RN metrics 
in the
previous section. 
If the RN black hole admits 
 an embeddeding
 in string theory (perhaps completed with other factors),
the throat region of the extremal metric with $M=Q$ 
 is a warped Scherk--Schwarz compactification  
with 
$$
L \,({\bf S}^1)_r = \beta\;\left(1-{Q\over r} \right) 
\,.
$$
Then, the correspondence radius $r_s$, defined in \corad, is nothing
but the ``stretched horizon" of Refs. \refs\rstretch.

This analogy suggests the identification of the stringy 
tachyonic instability
with the black-hole instability, and the corresponding flow with
the process of black-hole growth. The evidence from orbifold decay
\refs\raps\ supports this hypothesis, i.e. the ``ironing" of the
orbifold into flat space is topologically analogous to the ``capping"
of the throat by an Euclidean black-hole horizon \refs{\rusex, \rushag}\
 (c.f. Fig. 3).

 Based on this intuition, we can
derive a boundary condition on the stringy RG flows, giving the rough
features of the exit to the semiclassical regime. The proposal is
that the condensation amounts to an effective capping of the
non-contractible circle that supports the tachyons \refs{\rusex,\rushag}.
 When the capping
radius $r_0$ exceeds the correspondence radius $r_s$, the flow can
be approximated by a semiclassical flow of the type described in the
previous section.   The final stable geometry does not support winding
modes.  

What we propose
 is an off-shell version of the string/black-hole correspondence
principle. The general feature is the removal of that part of spacetime
that supports the tachyons topologically. 
In the following, we qualify our hypothesis  with a series
of remarks.

\subsec{Remarks}

{\bf (i)} 

The particular
 set of interpolating metrics with a conical singularity at
 $r=r_0$
is just one of many possible semiclassical 
flow trajectories that connect the unstable
solution of string theory at $r_0 = Q$ and the final non-extremal black-hole
background with Hawking temperature $1/\beta$. 
Along this particular trajectory, the world-sheet beta function receives
contributions  localized at the conical defect. One can imagine stringy
effects that smooth out the tip of the cone, 
perhaps along the lines of \refs\raps, as well as flows that excite the
dilaton.
Still, the  metrics with conical defect  give a good
 approximation  to the thermodynamic quantities (free energy, energy and
entropy) for $r_0 > r_s$.  

{\bf (ii)} 

The flow takes place in RG time, not in real time.
The Euclidean action (free energy)  is minimized along the flow and
thus it is similar to a thermal effective potential. Just as a thermal
effective potential compares the free energy of different {\it static}
 configurations
and has no direct relevance to the discussion of real time evolution
of the thermal instability, so the RG flow here must be distinguished
from real-time evolution (although we expect RG flow to give an approximation     
of a sufficiently adiabatic real-time flow). In principle,  real-time processes can
be realized in some cases as semiclassical ``bubbles of nothing".  
For example, one can consider asymptotic geometries of the form
 ${\bf S}^1 \times {\bf R}^{n+1}$ with
Minkowski signature and Scherk--Schwarz boundary conditions on the circle.
Then the analytic continuation of the Schwarzschild Euclidean section
yields the ``bubble of nothing", which is then interpreted as
a {\it vacuum} instability, rather than a thermal instability \refs{\rwitbub,
\rbmth, \rbath}. Unlike the  thermal
Euclidean flows discussed in this paper, the real-time
``bubbles of nothing" do not stabilize at finite distance and run away
to infinity.

{\bf (iii)} 

The decrease of the Euclidean effective action $I(X)$ along the
semiclassical flows is reminiscent of Zamolodchikov's monotonicity
in the central charge \refs{\rzamo,\ralud}. 
However, in the flows discussed above, the overall dimensionality of spacetime
 {\it does not} change. Rather, it is similar to  
 the case of boundary D-brane flows or tachyonic orbifold flows, i.e.  
the background  dynamics leaves a non-compact piece of spacetime essentially
untouched.  Thus, if the final background is semiclassical, its dimension
must coincide with the asymptotic dimension of the original unstable 
background.

Following  Ref. \refs\rpolyakov, one may try to  
 define a local notion of spacetime central charge
as
$
C(x) =  \CL_{\rm eff} (x)
$, 
 where $\CL_{\rm eff}$ denotes the integrand of $I(X)$. We assume  an additive
normalization so that $C(x) = D$ if the   initial unstable manifold had dimension
$D$ in the low-curvature limit. 
 Then, 
 in terms of the  coordinates $x^\mu$ that  parametrized the 
complete  unstable manifold $X$, one can say
that $C(x) \rightarrow 0$ on the region that has fallen behind the 
black-hole horizon.

 This concept of local central charge has some 
limitations. For example, it misses the importance of boundary contributions
to $I(X)$, the really relevant quantity.      
One indication of the primary relevance of the integrated action $I(X)$ comes
from holography.
 If $X$ has a holographic dual defined on $\partial X$, the classical action
$I(X)$ coincides with the quantum free energy of the dual in the planar
approximation. In this sense, 
 it is directly related to a counting of 
degrees of freedom.  

Notice that  
 the flow can be marginal at the classical level. This is the case of
orbifolds that are products of two-dimensional cones, as in \refs\raps.
In these cases
 the Einstein action
\ega\ is topological and invariant under
the ``ironing" of the orbifold. One must go to one-loop order  to see
the depletion of degrees of freedom through the effect of the 
``excluded volume". The corresponding one-loop density of states was
defined in \refs\rchicago.

{\bf (iv)}

Our proposal should be understood as a boundary condition for the  transition  
between stringy and  semiclassical parts of a given flow. Of course, we cannot
exclude the existence of other  
 endpoints (possibly metastable) lying entirely
 within the region of strong worldsheet coupling.  

A natural guess for such endpoints comes from considerations of T-duality. Since
it is the winding modes that condense, this corresponds to condensation of
non-zero momentum modes in T-dual language, a sort of ``spontaneous localization", 
such as that described in \refs{\rgkleb, \rforge}. Somewhat 
similar localization effects
can be seen in the case of flat orbifolds (c.f. \refs\rchicago).  All
these examples require special arrangements where symmetries highly constrain
the gravitational dynamics. This fact, together with the AdS/CFT considerations
of the next section, suggest that the black-hole like endpoint would be more
generic.

\newsec{Other Examples}

\noindent

In this section we consider two other examples of the main idea: off-shell
topology change as the semiclassical imprint of tachyon condensation processes
in string theory.

The first example is a generalization of the RN set up to the case of
D3-branes at finite temperature.  In this way we are able to present
a regularization of the Hagedorn behaviour of type IIB strings in
flat ten-dimensional spacetime. The second example is a semiclassical
picture  of the problem of D-brane-antiD-brane annihilation that   
illustrates the same general phenomenon.

\subsec{Hagedorn Behaviour in AdS Space}

\noindent

One example where the decoupling of the throat from the asymptotically 
flat spacetime is well understood is that of the AdS/CFT duality. 
The near-horizon region of $N$ D3-branes 
is ${\rm AdS}_5 \times {\bf S}^5$  with radius $R\sim \ell_s
(g_s N)^{1/4}$.  The same story that was told for the RN black hole can be
repeated here, with the advantage that one has a holographic interpretation
of the various actors. For example, the classical action $I(X)$ is
here interpreted as the true quantum free energy of the $\CN=4$ SYM theory
in the planar approximation. 

A further advantage of the AdS/CFT background is that it serves
as an example of a good adiabatic approximation. 
 Scherk--Schwarz compactifications with vanishing circles 
 can develop strong-coupling behaviour at
the singular locus. One argument for this is as follows: if $L({\bf S}^1)
\rightarrow 0$ is approached adiabatically, we may apply T-duality for
each circle at fixed $r$,  resulting in a 
 ``trumpet" like geometry  with diverging dilaton. Experience with
orbifolds of large deficit angle \refs\raps, together with the robustness
of black-hole thermodynamics, suggests that these effects will not
alter the general features of the endpoint proposed here.

    In the AdS/CFT example we can actually circumvent these subtleties by
considering the AdS space in global coordinates,  
\eqn\ads{
ds^2 = \left(1+{r^2 \over R^2} \right)\,d\tau^2 + \left(1+{r^2 \over R^2}
\right)^{-1} \,dr^2 + r^2 \,d\Omega^2
\;.} 
Picking thermal boundary conditions in Euclidean time $\tau \equiv \tau
+ \beta$, this background characterizes the  
 finite-temperature dual  CFT theory on a three-sphere ${\bf S}^3$ of
finite radius $R$. 

We have a warped compact circle with
$$
L\,({\bf S}^1)_r = \beta\,\sqrt{1+r^2 /R^2} \geq \beta 
\,.$$
Therefore, there is a lower bound on $L({\bf S}^1)$, attained at $r=0$, and
no singularity whatsoever. 
For $R\gg \ell_s$, the region $r<R$ is well approximated by the flat ${\bf S}^1 
\times {\bf R}^9$ background  
of type IIB string theory with thermal boundary conditions on the circle.  
Hagedorn tachyons are ``quasilocalized" in this region for $\beta < \ell_s$ and 
the adiabatic approximation is controlled by the $\alpha'$-corrections effective
parameter, $(g_s N)^{-1/2}$. 

According to the AdS/CFT dictionary, the thermal AdS metric \ads\  in the
Hagedorn regime,  $\beta < \ell_s \ll R$, 
would represent a highly superheated state of the dual CFT on  
   ${\bf S}^3$ of radius $R$, since the temperature is well above the 
Hawking--Page transition \refs{\rHPage, \rwithp}. 
One concludes that  this state is
highly unstable and will  thermalize back 
into the canonical equilibrium state. The instability is described in the gravity
side as the condensation of Hagedorn tachyons. The endpoint, on the other hand,
corresponds to the canonical equilibrium state of the high temperature CFT. In
the gravity dual it is described    
 by the large AdS black hole with horizon radius
$r_0 \sim R^2 /\beta$ and  Euclidean metric 
\eqn\adsbh{
ds^2 = \left(1+{r^2 \over R^2} - c {M \over r^2}\right)
\;d\tau^2 + \left(1+{r^2 \over R^2} - c {M \over r^2}\right)^{-1} \;dr^2 
+ r^2 \;d\Omega^2
\;,} 
with $c=16\pi/3 {\rm Vol}({\bf S}^3)$ and
\eqn\admm{
M={3{\rm Vol}({\bf S}^3) \over 16\pi } \left({r_0^4 \over R^2} +r_0^2 \right). 
}
Hence, AdS/CFT gives strong arguments in favour of our geometrical interpretation
of the flow \refs\rushag. 

Just as before, we can define a matching radius $r_s$ by the requirement that
the thermal circle be stringy:
\eqn\match{
\beta \,\sqrt{1+ r_s^2 /R^2} = \ell_s
\,.} 
In this case, we see that $r_s \sim R$ for $\beta < \ell_s$ unless $\beta/\ell_s$
is fine-tuned to unity.
 The conjecture is that the stringy part of the tachyon condensation
on ${\bf S}^1 \times {\bf R}^9$ ends up with the depletion of the whole flat
spacetime and one has an ${\rm AdS}_5 \times {\bf S}^5$ metric capped at 
$r \sim r_s \sim R$. After this comes
the semiclassical part of the flow. It corresponds to  
black holes that are smaller than the stable AdS black-hole and grow because
they are cooler than the asymptotic thermal boundary conditions. Such off-shell
black holes can be modelled by the metric \adsbh\ with fixed $\beta$ and
varying $r_s < r_0 < R^2 /\beta$,
 and a conical sigularity at the horizon.

\fig{\sl The $d\Omega =0$ section of the 
 thermal AdS manifold. It has a cylindrical part that is unstable under
condensation of string winding modes, as shown here.
 The endpoint of the decay process must be the Euclidean large AdS
black hole (in dark), which does not support string
 winding modes. In the $R/\beta \rightarrow \infty$
limit   the unstable cylinder  approximates ${\bf S}^1 \times {\bf
R}^{9}$ whereas the stable endpoint metric recedes to infinity and approximates
locally ${\bf R}^{10}$.  
 We also show in dashed lines  an interpolating
off-shell black-hole metric with a conical singularity at the horizon.   
}{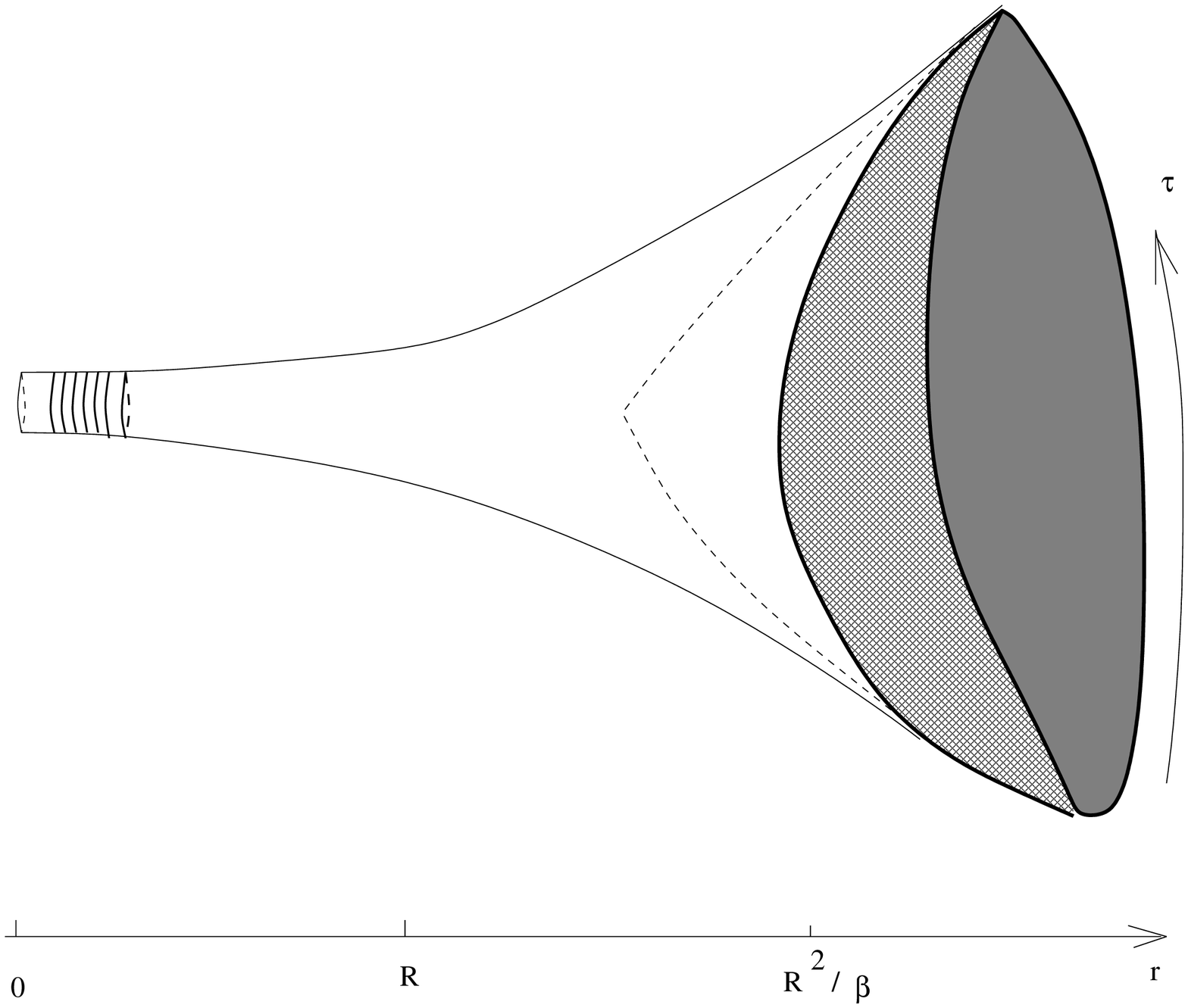}{5truein}

The local curvature of the final large AdS black-hole geometry is of order
$(g_s N)^{-1/2}$ in string units. Hence, it vanishes in the flat-space limit 
$R/\ell_s \rightarrow \infty$ in which we remove the infrared AdS regulator
of the ${\bf S}^1 \times {\bf R}^9$ background. In this sense, we can say that
the endpoint of tachyon condensation of ${\bf S}^1 \times {\bf R}^9$ 
 is the  flat ${\bf R}^{10}$ Euclidean background of type IIB string theory. 
Thus, we can make contact with similar proposals in the literature that
connect various processes of tachyon condensation with supersymmetric endpoints
 \refs{\rbunch, \raps,  \rgutperlem}.   
Our picture gives a nice infrared regularization of the process 
 in which the flat endpoint background sits on a ``transverse" plane in ten
dimensions (see \refs\rushag\ for details).
 This is entirely analogous to the decay of nonsupersymmetric
cones into flat space as described in \refs\raps.

There is an interesting connection between this scenario of  
 Hagedorn behaviour and that of   Ref. \refs\raw, where it was
argued   
that the effective potential for the thermal tachyon on ${\bf S}^1
\times {\bf R}^9$ would naturally yield a first-order phase transition
at a temperature slightly below the nominal Hagedorn temperature. 
In our picture, we identify the expectation value of the thermal tachyon
with the extent of the ${\bf S}^1$ capping at the correspondence point.
Therefore, it is natural to identify 
the Atick--Witten effective potential itself with the 
Euclidean gravitational action, $I(r_0)$, 
 as a function of the capping radius $r_0$  
$$
\beta\,{\rm Vol} \,({\bf R}^9)\,V_{\rm eff} \approx I(r_0)
\,.$$  
In this equation, the volume of ${\bf R}^{9}$ should be regularized to
be of order $R^9$.  

To see the first-order phase transition, notice that
  the
action of AdS black holes, 
$$
I(r_0) = \beta\,M(r_0)- S(r_0) = {3\,{\rm Vol} ({\bf S}^3) \over 16\pi}
\,\left[{\beta \,r_0^4 \over R^2 } + \beta \,r_0^2 - {4 \,r_0^3 \over 3} \right]
\,,$$  
 has two extrema for
$\beta \gg \ell_s$. One is the
large stable AdS black hole with $r_0 \sim R^2 /\beta$, and the other
one is the small (unstable) AdS black hole with $r_0 \sim \beta$. The small
AdS black hole is a local maximum of $I(r_0)$ and the value of
the action there is positive.  Thus,
for temperatures slightly below the Hagedorn temperature, the transition
from the hot AdS space to the large AdS black hole 
proceeds initially by tunneling through a barrier, just like in a
first order phase transition (of course, there is also the Jeans' instability,
but this occurs on a larger length scale).   

The transition amplitude was actually calculated 
in \refs\rgpy, in the case of Schwarzschild black holes in flat space.
In the  AdS spacetime the  tunneling rate can be
estimated as  
$$
\Gamma \sim e^{-I(r_0)} \sim \exp\left[-{C \over g_s^2} {\beta^3 R^5 \over 
\ell_s^8}\right]
\,.$$
Small ten-dimensional
Schwarzschild black holes have a rate
$$
\Gamma \sim \exp\left[-{C' \over g_s^2} {\beta^8 \over \ell_s^8}\right]
\,,$$
which actually dominates for $\beta < R$. 
On closer scrutiny, one finds that these rates are still non-perturbatively
suppressed, $\Gamma \sim \exp(-1/g_s^2)$ for $\beta \sim \ell_s$. However,
we cannot trust the WKB estimation to this extent. To see this, notice that the
variation of the Euclidean action  around the maximum is of order
 $\Delta I \sim g_s^{-2}$ 
when the horizon varies on the
string scale $\Delta r_0 \sim \ell_s$. This means that  the low-energy approximation
breaks down in the calculation of the barrier when $\beta \sim \ell_s$.
 A natural expectation
at this point is that the barrier is completely washed out by 
$\alpha'$ corrections, so that the tachyon condensation proceeds without 
any barriers when $\beta < \ell_s$, i.e. it becomes a classical process.

One interesting aspect of this scenario is that it potentially
reconciles the (historically orthogonal)
 microcanonical and canonical pictures of the  Hagedorn transition.
The essential ingredient in the microcanonical picture of the
 AdS Hagedorn transition is the string/black-hole
correspondence principle \refs\rHP.
 In a sense, we have just incorporated this correspondence
principle
into the Euclidean canonical picture.

\subsec{Open-String Tachyons and Topology Change}

\noindent

An interesting testing ground for these ideas is the much better understood 
condensation of  open-string tachyons 
in the ${\rm D}p-{\overline{\rm D}}p$ system. States of stretched
strings with ground-state mass
\eqn\spec{
m^2 = \left({d_\perp \over 2\pi\alpha'} \right)^2 - {c \over \alpha'} 
}
become tachyonic for $d_\perp < \CO(\ell_s)$, representing the instability
towards brane-antibrane annihilation.
For $d_\perp \gg \ell_s$ the D-branes are essentially stable, except for
the small gravitational attraction. They are  also non-perturbatively
unstable: if they pay the energy necessary to reconnect via a
``wormhole" (by virtual
tunneling), then
they can subsequently lower their energy by the wormhole
 running off to infinity.

The basic tools to study this process can be found in \refs\rcmalda. Using
the non-linear Dirac--Born--Infeld action, one can derive explicit
expressions for the wormholes. For a pair of parallel branes separated a distance
$d_\perp$, one finds two types of wormholes. There is a large unstable solution
with a neck of size $r_0 \sim d_\perp$ and energy (normalized to the bare
tension
of the parallel branes):
$$
E_{\rm large} \sim {1\over g_s \ell_s} \left({r_0 \over \ell_s}\right)^p \sim
{1\over g_s \ell_s} \left({d_\perp \over \ell_s}\right)^p
\,.$$
(In the derivation of this formula it is actually needed to assume $p\geq 3$).
The semiclassical nucleation of such bubbles has rate
$$
\Gamma \sim \exp\,\left[-{C \over g_s} \left({d_\perp \over \ell_s}\right)^{
p+1} \right]
\;.$$
Keeping the string coupling small, it is natural to assume that this
process matches the standard open-string tachyon roll when $d_\perp \sim
\ell_s$. Again, the WKB barrier must be smoothed out by $\alpha'$ corrections.

The interesting fact about this example is that it makes this correspondence
principle basically inescapable. Consider the case of rotated branes, say
by a $\pi/2$ angle, in such a way that the $c$ appearing in \spec\ is positive
and  there are tachyons for $d_\perp < \CO(\ell_s)$. Now the tachyonic
open strings are supported {\it only} in the vicinity of the ``intersection"
(the region of the branes within $\CO(d_\perp)$ distance). We know that
whatever the microscopic description of the decay would be, at very large
distances the decay proceeds by reconnection of the branes that subsequently
run off to infinity, i.e. by the analogue of the wormhole process (c.f.
 Figure 13.4 in \refs\rpolbook). Therefore,
there must be a matching between the boundary RG flow that describes the
open-string tachyon condensation, and the run-away of the reconnected branes.  

The small wormhole solution turns out to be the solitionic description
of the fundamental open string stretching between 
the D-brane pair. 
This  is a bion with neck size $r_0$ given by
$(r_0/\ell_s)^{p-2}  \sim
g_s \ell_s /d_\perp$. It is locally stable  for large $d_\perp$ and has
energy
$$
E_{\rm small} \sim {d_\perp \over \ell_s^2}
$$
as expected for the stretched fundamental string. When $d_\perp$
is lowered to the critical value
$$
(d_\perp)_{\rm critical}  \sim \ell_s \;g_s^{\;\;1\over  p-1}
$$
the two solutions coalesce and all wormholes of any neck size become
unstable towards growth! Hence, it is tempting to regard this process of
``blow up" of the tachyonic fundamental string as the geometrical
interpretation of the tachyon condensation process. Because the system
is non-supersymmetric, the matching length scale could receive $\CO(1)$
corrections in string units, so that  the true instability should take
place at $d_\perp \sim \ell_s$.

We see that open-string tachyon condensation, a much better understood
phenomenon, is also compatible with our principle of off-shell topology
change. The analogue of the ``cylinder capping" in this case is the 
formation of the wormhole between the two parallel D-branes. The tachyonic
modes are topologically supported, since they are just the
 fundamental strings stretched between the two D-branes.  
Once the wormhole is formed, a stretched string in the far region can
be ``unwrapped" through the wormhole neck.

\newsec{An Off-Shell Correspondence Principle?}

\noindent

We can abstract from the previous examples  the following rule: if the
condensation of topologically-supported tachyons has a semiclassical
regime, the geometrical interpretation of this regime involves a dynamical
topology-change process in the background. The background dynamics consists  
primarily on the formation of ``holes" or ``bubbles" in such a way that the
topologically supported states can be unwrapped.  

In the remainder of this section we speculate on a more quantitative
version of this principle. 
Nonperturbative instabilities in string theory that can be studied
in the semiclassical approximation have rates of the form
$$
\exp \; \left[-I(\ell_{\rm eff})\right] = 
\exp\;\left[-{C\over g_s^a} \;\left({\ell_{\rm eff}
 \over \ell_s}\right)^b \right] 
\;,
$$
where
$\ell_{\rm eff}$ is a characteristic length scale of the decay process.
This formula covers the previously considered cases of closed-string
instabilities, with $a=2$, and also cases of open-string instabilities, for
which $a=1$. 
The exponent $b$ is related to 
the effective dimensionality on which the process takes place. 

A natural  conjecture at this point would be that
all these processes match tree-level instabilities (tachyon instabilities)
as $\ell_{\rm eff} \sim \ell_s$, in a dynamical generalization of the
string/black-hole correspondence principle. For this matching to make
sense one has to assume two properties. First, $\alpha'$ corrections should
wash out the WKB barrier at the matching point. Second, we should ensure
that the dilaton is small throughout the background in order to apply
the classical approximation to the estimate of the decay rate.

It has been proposed  that the semiclassical regime and the
string tachyon condensation should be dually related. The simplest
idea is that some semiclassical decay process may become a tachyon
condensation process when the theory is continued to strong coupling
$g_s \rightarrow \infty$ so that the nonperturbative exponent becomes
small (the barrier collapses).
 Characteristic examples of such conjectures 
are bosonic M-theory and the interpretation of various tachyonic
 Type 0 models via duality to Type II Scherk--Schwarz 
compactifications \refs{\rbmth, \rgutperlem}. 
These proposals  
 involve strong-coupling extrapolations of processes that are not
protected by any supersymmetry and change radically the properties
of the asymptotic vacuum. 

We propose here a weaker version of the correspondence between
semiclassical decay processes and tachyon condensation. 
 In this version one mimics the correspondence principle between black holes
and strings. In that case, one matches {\it finite energy} configurations
as a function of some control parameter, but one stays at weak string
coupling, so that the asymptotic vacuum is described by the same string
theory on both sides of the correspondence. In other words, it is
a change in description from microscopic to semiclassical for a class
of states of the Hilbert space, but the vacuum does not change in
the process.

Analogously, we propose a matching between the semiclassical effective
action $I(\ell_{\rm eff})$ and the corresponding effective potential of the 
microscopic RG flow, without any change of the asymptotic boundary conditions,
i.e. with no change of asymptotic vacuum, which is supposed weakly coupled.
We think that this principle, although somewhat qualitative, is useful because
it seems to capture the essential geometry of tachyonic instabilities in
string theory, as we have argued over several examples of  
very different nature.

\vskip0.2cm

{\bf Acknowledgments}

We would like to thank Roberto Emparan, Barak Kol and Miguel A. V\'azquez-Mozo for
discussions. 
The work of J.L.F.B. was partially supported by MCyT 
 and FEDER under grant
BFM2002-03881 and  
 the European RTN network 
 HPRN-CT-2002-00325. The work of E.R. is supported in part by the
BSF-American Israeli Bi-National Science Foundation, The Israel Science
Foundation-Centers of Excellence Program, The German-Israel Bi-National
Science Foundation and the European RTN network HPRN-CT-2000-00122.  

\listrefs

%\vfill\eject
\bye